\documentclass[12pt]{article}
\usepackage{pdproc,epsfig} 

\def\dd{\displaystyle}
\def\nn{\nonumber}
\def\bea{\begin{eqnarray}}
\def\eea{\end{eqnarray}}
\def\beq{\begin{equation}}
\def\eeq{\end{equation}}
\def\bq{\begin{quote}}
\def\eq{\end{quote}}
\def\gappeq{\mathrel{\rlap {\raise.5ex\hbox{$>$}} {\lower.5ex\hbox{$\sim$}}}}
\def\lappeq{\mathrel{\rlap{\raise.5ex\hbox{$<$}} {\lower.5ex\hbox{$\sim$}}}}

\def\be{\begin{equation}}
\def\ee{\end{equation}}
\def\bc{\begin{center}}
\def\ec{\end{center}}
\def\bea{\begin{eqnarray}}
\def\eea{\end{eqnarray}}
\def\dd{\displaystyle}
\def\nn{\nonumber}
\def\gappeq{\mathrel{\rlap {\raise.5ex\hbox{$>$}} {\lower.5ex\hbox{$\sim$}}}}
\def\lappeq{\mathrel{\rlap{\raise.5ex\hbox{$<$}} {\lower.5ex\hbox{$\sim$}}}}
\newcommand{\bac}{\beq\begin{array}}
\newcommand{\eac}{\end{array}\eeq}
\newcommand{\ba}{\begin{array}}
\newcommand{\ea}{\end{array}}
\newcommand{\beaa}{\begin{eqnarray*}}
\newcommand{\eeaa}{\end{eqnarray*}}
\newcommand{\La}{\Lambda}
\newcommand{\mean}[1]{\langle#1\rangle}
\begin{document}
  
\vspace*{-1cm}
\phantom{hep-ph/***}

\hfill{RM3-TH/09-9}
\hfill{CERN-PH-TH/2009-060}

\vskip 2.5cm

\renewcommand{\thefootnote}{\alph{footnote}}
  
\title{THEORETICAL MODELS OF NEUTRINO MIXING:\\ 
RECENT DEVELOPMENTS}

\author{GUIDO ALTARELLI}

\address{Dipartimento di Fisica, Universita' di Roma Tre\\
Rome, Italy\\
and\\
CERN, Department of Physics, Theory Division \\  
CH-1211 Gen\`eve 23, Switzerland\\
{\rm E-mail: guido.altarelli@cern.ch}}

\abstract{The data on neutrino mixing are at present compatible with Tri-Bimaximal (TB) mixing. If one takes this indication seriously then the models that lead to TB mixing in first approximation are particularly interesting and $A_4$ models are prominent in this list. However, the agreement of TB mixing with the data could still be an accident. We discuss a recent model based on $S_4$ where Bimaximal mixing is instead valid at leading order and the large corrections needed to reproduce the data arise from the diagonalization of charged leptons. The value of $\theta_{13}$ could distinguish between the two alternatives}

\normalsize\baselineskip=15pt

\vskip 2cm

\section{Introduction}

It is an experimental fact \cite{data,FogliIndication,MaltoniIndication} that within measurement errors
the observed neutrino mixing matrix \cite{review} is compatible with
the so called Tri-Bimaximal (TB) form \cite{hps}. The best measured neutrino mixing angle $\theta_{12}$ is just about 1$\sigma$ below the TB value $\tan^2{\theta_{12}}=1/2$, while the other two angles are well inside the 1$\sigma$ interval (see table \ref{table:OscillationData}).

\begin{table}[h]
\begin{center}
\begin{tabular}{|c|c|c|}
\hline
&&\\[-4mm]
  & ref. \cite{FogliIndication} & ref. \cite{MaltoniIndication}   \\[2mm]
\hline
&&\\[-4mm]
$\sin^2\theta_{12}$ &$0.326^{+0.050}_{-0.040}~~~[2\sigma]$ & $0.304^{+0.022}_{-0.016}$\\[2mm]
\hline
&&\\[-4mm]
$\sin^2\theta_{23}$ &$0.45^{+0.16}_{-0.09}~~~[2\sigma]$ &  $0.50^{+0.07}_{-0.06}$\\[2mm]
\hline
&&\\[-4mm]
$\sin^2\theta_{13}$ &$0.016\pm0.010$ &$0.010^{+0.016}_{-0.011}$ \\[2mm]
\hline
  \end{tabular}
\end{center}
\begin{center}
\begin{minipage}[t]{12cm}
\caption{\label{table:OscillationData}Results of two recent fits to the lepton mixing angles.}
\end{minipage}
\end{center}
\end{table}

Thus, one possibility is that one takes this coincidence seriously and considers models where TB mixing is a good first approximation. Alternatively one can assume that this agreement of the data with TB mixing is accidental. Indeed there are many models that fit the data and yet TB mixing does not play a role in their architecture. For example, in ref.(\cite{alro}) there is a list of Grand Unified SO(10) models with fits to the  neutrino mixing angles that show good agreement with the data although most of them  have no relation with TB mixing. Another class of examples is found in ref.(\cite{seidl}). However, in most cases, for this type of models different mixing angles could also be accommodated by simply varying the fitted values of the parameters. If instead we assume that TB mixing has a real dynamical meaning then it is important to consider models that naturally lead to TB mixing. In a series of papers \cite{TBA4,AFextra,AFmodular,AFL,afh,altverlin,altveram} it has been pointed out that a broken flavour symmetry based on the discrete
group $A_4$ appears to be particularly suitable to reproduce this specific mixing pattern in Leading Order (LO). Other
solutions based on alternative discrete or  continuous flavour groups have also been considered \cite{continuous,others,bmm}, but the $A_4$ models have a very economical and attractive structure, e.g. in terms of group representations and of field content. 
In most of the models $A_4$ is accompanied by additional flavour symmetries, either discrete like $Z_N$ or continuous like U(1), which are necessary to eliminate unwanted couplings, to ensure the needed vacuum alignment and to reproduce the observed mass hierarchies. Given the set of flavour symmetries and having specified the field content, the non leading corrections to the TB mixing arising from higher dimensional effective operators can be evaluated in a well defined expansion. In the absence of specific dynamical tricks, in a generic model, all the three mixing angles receive corrections of the same order of magnitude. Since the experimentally allowed departures of $\theta_{12}$ from the TB value $\sin^2{\theta_{12}}=1/3$ are small, at most of $\mathcal{O}(\lambda_C^2)$, with $\lambda_C$ the Cabibbo angle, it follows that both $\theta_{13}$ and the deviation of $\theta_{23}$ from the maximal value are expected in these models to also be at most of $\mathcal{O}(\lambda_C^2)$ (note that $\lambda_C$ is a convenient hierarchy parameter not only for quarks but also in the charged lepton sector with $m_\mu/m_\tau \sim0.06 \sim \lambda_C^2$ and $m_e/m_\mu \sim 0.005\sim\lambda_C^{3-4}$). A value of $\theta_{13} \sim \mathcal{O}(\lambda_C^2)$ is within the sensitivity of the experiments which are now in preparation and will take data in the near future. 

Going back to the possibility that the agreement of the data with TB mixing is accidental, we observe that the present data do not exclude a larger value for $\theta_{13}$,$\theta_{13} \sim \mathcal{O}(\lambda_C)$, than generally implied by models with approximate TB mixing. In fact, two recent analysis of the available data lead to
$\sin^2{\theta_{13}}=0.016\pm0.010$ at 1$\sigma$ \cite{FogliIndication} and $\sin^2{\theta_{13}}=0.010^{+0.016}_{-0.011}$ at 1$\sigma$ \cite{MaltoniIndication}, which are compatible with both options. If experimentally it is found that $\theta_{13}$ is near its present upper bound, this could be interpreted as an indication that the agreement with the TB mixing is accidental. Then a scheme where instead the Bimaximal (BM) mixing is the correct first approximation modified by terms of $\mathcal{O}(\lambda_C)$ could be relevant. This is in line with the well known empirical observation that $\theta_{12}+\lambda_C\sim \pi/4$, a relation known as quark-lepton complementarity \cite{compl}, or similarly $\theta_{12}+\sqrt{m_\mu/m_\tau} \sim \pi/4$. No compelling model leading, without parameter fixing, to the exact complementarity relation has been produced so far. Probably the exact complementarity relation is to be replaced with something like $\theta_{12}+\mathcal{O}(\lambda_C)\sim \pi/4$ or $\theta_{12}+\mathcal{O}(m_\mu/m_\tau)\sim \pi/4$ (which we could call "weak" complementarity).
Along this line of thought, we have used the expertise acquired with non Abelian finite flavour groups to construct a model \cite{S4us} based on the permutation group $S_4$ which naturally leads to the BM mixing at LO. We have adopted a supersymmetric formulation of the model in 4 space-time dimensions. The complete flavour group is $S_4\times Z_4 \times U(1)_{FN}$. In LO, the charged leptons are diagonal and hierarchical and the light neutrino mass matrix, after see-saw, leads to the exact BM mixing. The model is built in such a way that the dominant corrections to the BM mixing, from higher dimensional operators in the superpotential,  only arise from the charged lepton sector at Next-to-the-Leading-Order (NLO)  and naturally inherit $\lambda_C$ as the relevant expansion parameter. As a result the mixing angles deviate from the BM values by terms of  $\mathcal{O}(\lambda_C)$ (at most), and weak complementarity holds. A crucial feature of the model is that only $\theta_{12}$ and $\theta_{13}$ are corrected by terms of $\mathcal{O}(\lambda_C)$ while $\theta_{23}$ is unchanged at this order (which is essential to make the model agree with the present data). 

In this concise review of recent developments I will first make an update on $A_4$, also discussing our Grand Unified model with TB mixing in the lepton sector from $A_4$ \cite{afh}, and then present the alternative possibility of BM mixing from $S_4$ corrected by terms of  $\mathcal{O}(\lambda_C)$ from the charged lepton sector.

\section{Tri-Bimaximal Mixing and $A_4$}

The  TB mixing matrix (in a particular phase convention) is given by:
\begin{equation}
U_{TB}= \left(\matrix{
\dd\sqrt{\frac{2}{3}}&\dd\frac{1}{\sqrt 3}&0\cr
-\dd\frac{1}{\sqrt 6}&\dd\frac{1}{\sqrt 3}&-\dd\frac{1}{\sqrt 2}\cr
-\dd\frac{1}{\sqrt 6}&\dd\frac{1}{\sqrt 3}&\dd\frac{1}{\sqrt 2}}\right)~~~~~. 
\label{2}
\end{equation}
As we have already mentioned this matrix is supported by the present data within a $1-\sigma$ accuracy (see table \ref{table:OscillationData}).

The TB mixing matrix suggests that mixing angles are independent of mass ratios (while for quark mixings relations like $\lambda_C^2\sim m_d/m_s$ are typical). In fact in the basis where charged lepton masses are 
diagonal, the effective neutrino mass matrix in the TB case is given by $m_{\nu}=U_{TB}\rm{diag}(m_1,m_2,m_3)U_{TB}^T$:
\begin{equation}
m_{\nu}=  \left[\frac{m_3}{2}M_3+\frac{m_2}{3}M_2+\frac{m_1}{6}M_1\right]~~~~~. 
\label{1k1}
\end{equation}
where:
\be
M_3=\left(\matrix{
0&0&0\cr
0&1&-1\cr
0&-1&1}\right),~~~~~
M_2=\left(\matrix{
1&1&1\cr
1&1&1\cr
1&1&1}\right),~~~~~
M_1=\left(\matrix{
4&-2&-2\cr
-2&1&1\cr
-2&1&1}\right).
\label{4k1}
\ee
The eigenvalues of $m_{\nu}$ are $m_1$, $m_2$, $m_3$ with eigenvectors $(-2,1,1)/\sqrt{6}$, $(1,1,1)/\sqrt{3}$ and $(0,1,-1)/\sqrt{2}$, respectively. The expression in eq.(\ref{1k1}) can be reproduced in models with sequential dominance or with form dominance, discussed by S. King and collaborators \cite{ski}. 

As we see the most general neutrino mass matrix corresponding to TB mixing, in the basis of diagonal charged leptons, is of the form:
\begin{equation}
m=\left(\matrix{
x&y&y\cr
y&x+v&y-v\cr
y&y-v&x+v}\right),
\label{gl21}
\end{equation}
This is a symmetric, 2-3 symmetric matrix with $a_{11}+a_{12}=a_{22}+a_{23}$.

We recall that $A_4$, the group of even permutations of 4 objects, can be generated by the two elements
$S$ and $T$ obeying the relations (a "presentation" of the group):
\be
S^2=(ST)^3=T^3=1~~~.
\label{$A_4$}
\ee
The 12 elements of $A_4$  are obtained as:
$1$, $S$, $T$, $ST$, $TS$, $T^2$, $ST^2$, $STS$, $TST$, $T^2S$, $TST^2$, $T^2ST$.
The inequivalent irreducible representations of $A_4$ are 1, 1', 1" and 3. It is immediate to see that one-dimensional unitary representations are
given by:
\be
\begin{array}{lll}
1&S=1&T=1\\
1'&S=1&T=e^{\dd i 4 \pi/3}\equiv\omega^2\\
1''&S=1&T=e^{\dd i 2\pi/3}\equiv\omega\\
\label{s$A_4$}
\end{array}
\ee
The three-dimensional unitary representation, in a basis
where the element $T$ is diagonal, is given by:
\be
T=\left(
\begin{array}{ccc}
1&0&0\\
0&\omega^2&0\\
0&0&\omega
\end{array}
\right),~~~~~~~~~~~~~~~~
S=\frac{1}{3}
\left(
\begin{array}{ccc}
-1&2&2\cr
2&-1&2\cr
2&2&-1
\end{array}
\right)~~~.
\label{ST}
\ee
It is useful to remind the product rules of two triplets, ($\psi_1,\psi_2,\psi_3$) and ($\varphi_1,\varphi_2,\varphi_3$) of $A_4$, according to the multiplication rule 3x3=1+1'+1"+3+3:
\bea \label{tensorproda4}
&\psi_1\varphi_1+\psi_2\varphi_3+\psi_3\varphi_2 \sim 1 ~,\nn \\
&\psi_3\varphi_3+\psi_1\varphi_2+\psi_2\varphi_1 \sim 1' ~,\nn \\
&\psi_2\varphi_2+\psi_3\varphi_1+\psi_1\varphi_3 \sim 1'' ~,\nn
\eea
  \be
   \left( 
 \ba{c}
2\psi_1\varphi_1-\psi_2\varphi_3-\psi_3\varphi_2 \\
2\psi_3\varphi_3-\psi_1\varphi_2-\psi_2\varphi_1 \\
2\psi_2\varphi_2-\psi_1\varphi_3-\psi_3\varphi_1 \\
  \ea
  \right) \sim 3_S~, \qquad
  \left( 
 \ba{c}
\psi_2\varphi_3-\psi_3\varphi_2 \\
\psi_1\varphi_2-\psi_2\varphi_1 \\
\psi_3\varphi_1-\psi_1\varphi_3 \\
  \ea
  \right) \sim 3_A~.
  \label{tensorp}
   \ee
   
Note that the most general mass matrix for TB mixing in eq.(\ref{gl21}) can be specified as the most general matrix which is invariant under $\mu-\tau$ symmetry and under the $S$ unitary transformation:
\bea
m=SmS,~~~~~m=A_{\mu \tau}mA_{\mu \tau}~~\label{inv}
\eea
where S is given in eq.(\ref{ST}) and:
\be
A_{\mu \tau}=\left(
\begin{array}{ccc}
1&0&0\\
0&0&1\\
0&1&0
\end{array}
\right)
\label{Amutau}
\ee
This observation plays a role in leading to $A_4$ as a candidate group for TB mixing, because $S$ is a matrix of $A_4$ (but  $A_{\mu \tau}$ is not).

In the lepton sector a typical A4 model works as follows \cite{AFmodular} (many alternative versions can be found in the literature; in particular recently some models were proposed with a different alignment such that the charged lepton hierarchies are obtained without introducing a $U(1)$ symmetry \cite{altverlin,altveram}).  One assigns
leptons to the four inequivalent
representations of A4: left-handed lepton doublets $l$ transform
as a triplet $3$, while the right-handed charged leptons $e^c$,
$\mu^c$ and $\tau^c$ transform as $1$, $1''$ and $1'$, respectively. 
Here we consider a  see-saw realization, so we also introduce conjugate right-handed neutrino fields $\nu^c$ transforming as a triplet of $A_4$. We adopt a supersymmetric context also to make contact with Grand Unification and the models of the following sections. The flavour symmetry is broken by two triplets
$\varphi_S$ and $\varphi_T$ and by a singlet $\xi$. Actually we may need more singlets but in this section we only keep the terms with $\xi$ for simplicity.
All these fields are gauge singlets.
Two Higgs doublets $h_{u,d}$, invariant under $A_4$, are
also introduced. One can obtain  the observed hierarchy among $m_e$, $m_\mu$ and
$m_\tau$ by introducing an additional U(1)$_{FN}$ flavour symmetry under
which only the right-handed lepton sector is charged.
We assign FN-charges $0$, $2$ and $4$ to $\tau^c$, $\mu^c$ and
$e^c$, respectively. By assuming that a flavon $\theta$, carrying
a negative unit of FN charge, acquires a VEV 
$\langle \theta \rangle/\Lambda\equiv\lambda<1$, the Yukawa couplings
become field dependent quantities $y_{e,\mu,\tau}=y_{e,\mu,\tau}(\theta)$
and we have
\be
y_\tau\approx O(1)~~~,~~~~~~~y_\mu\approx O(\lambda^2)~~~,
~~~~~~~y_e\approx O(\lambda^{4})~~~.
\ee

The superpotential term for lepton masses, $w_l$ is given by:
\be
w_l=y_e e^c (\varphi_T l)+y_\mu \mu^c (\varphi_T l)'+
y_\tau \tau^c (\varphi_T l)''+ y (\nu^c l)+
(x_A\xi+\tilde{x}_A\tilde{\xi}) (\nu^c\nu^c)+x_B (\varphi_S \nu^c\nu^c)+h.c.+...
\label{wlss}
\ee
dots denoting higher-order contributions. In our notation, $(3 3)$ transforms as $1$, 
$(3 3)'$ transforms as $1'$ and $(3 3)''$ transforms as $1''$. 
To keep our formulae compact, we omit to write the Higgs and flavon  fields
$h_{u,d}$, $\theta$ and the cut-off scale $\Lambda$. For instance 
$y_e e^c (\varphi_T l)$ stands for $y_e e^c (\varphi_T l) h_d \theta^4/\Lambda^5$. Dots stand for higher
dimensional operators that will be discussed later on. 
Some terms allowed by the flavour symmetry, such as the terms 
obtained by the exchange $\varphi_T\leftrightarrow \varphi_S$, 
(or the term $(\nu^c\nu^c)$) are missing in $w_l$. 
Their absence is crucial and, in each version of A4 models, is
motivated by additional symmetries. Here the additional symmetry is $Z_3$. A $U(1)_R$ symmetry related to R-parity and the presence of driving fields in the flavon superpotential are common features of supersymmetric formulations. Supersymmetry also helps producing and maintaining the hierarchy $\langle h_{u,d}\rangle=v_{u,d}\ll \Lambda$ where $\Lambda$ is the cut-off scale of the theory.
The fields in the model and their classification under the symmetry are summarized in Table \ref{table:TransformationsA}.

\begin{table}[h]
\begin{center}
\begin{tabular}{|c||c|c|c|c|c|c||c|c|c|c||c|c|c|}
  \hline
  &&&&&&&&&&&&&\\[-0,3cm]
  & $l$ & $e^c$ & $\mu^c$ & $\tau^c$ & $\nu^c$ & $h_{u,d}$ & $\theta$ & $\varphi_T$ & $\varphi_S$ &  $\xi$ &$\varphi_0^T$ & $\varphi_0^S$ &  $\xi_0$ \\
 &&&&&&&&&&&&&\\[-0,3cm]
  \hline
  &&&&&&&&&&&&&\\[-0,3cm]
  $A_4$ & 3 & 1 & 1" & 1' & 3 & 1 & 1 & 3 & $3$ & 1 & 3 & 3 & 1  \\
   &&&&&&&&&&&&&\\[-0,3cm]
  $Z_3$ & $\omega$ &$\omega^2$ & $\omega^2$& $\omega^2$& $\omega^2$ & 1 & 1 &1& $\omega^2$&$\omega^2$& 1 &$\omega^2$ & $\omega^2$  \\
   &&&&&&&&&&&&&\\[-0,3cm]
  $U(1)_{FN}$ & 0 & 4 & 2 & 0 & 0 & 0 & -1 & 0 & 0 & 0 & 0 & 0 & 0  \\
  &&&&&&&&&&&&&\\[-0,3cm]
  $U(1)_R$ & 1 & 1 & 1 & 1 & 1 & 0 & 0 & 0 & 0 & 0 & 2 & 2 & 2  \\
  \hline
  \end{tabular}
\end{center}
\begin{center}
\begin{minipage}[t]{12cm}
\caption[]{\label{table:TransformationsA}Transformation properties of all the fields.}
\end{minipage}
\end{center}
\end{table}
In this set up it can be shown that the fields $\varphi_T$,
$\varphi_S$ and $\xi$ develop a VEV along the directions:
\bea
\langle \varphi_T \rangle&=&(v_T,0,0)\nn\\ 
\langle \varphi_S\rangle&=&(v_S,v_S,v_S)\nn\\
\langle \xi \rangle&=&u~~~. 
\label{align}
\eea 
A crucial part of all serious A4 models is the dynamical generation of this alignment in a natural way. We refer to ref. \cite{AFmodular} for a proof that the above alignment naturally follows from the most general LO superpotential implied by the symmetries of the model. The group $A_4$ has two obvious subgroups: $G_S$, which is a reflection subgroup
generated by $S$ and $G_T$, which is the group generated by $T$, isomorphic to $Z_3$.
In the basis where $S$ and $T$ are given by eq.(\ref{ST}), the VEV $\langle \varphi_T \rangle=(v_T,0,0)$ breaks $A_4$ down to $G_T$, while
$\langle \varphi_S\rangle=(v_S,v_S,v_S)$
breaks $A_4$ down to $G_S$.  

If the alignment in eq.(\ref{align}) is realized, at the leading order of the $1/\Lambda$ expansion,
the mass matrices $m_l$ and $m_\nu$ for charged leptons and 
neutrinos can be derived. The charged lepton mass matrix is diagonal:
\be
m_l=v_d\frac{v_T}{\Lambda}\left(
\begin{array}{ccc}
y_e& 0& 0\\
0& y_\mu & 0 \\
0& 0& y_\tau 
\end{array}
\right)~~~,
\label{mch}
\ee
The charged fermion masses are given by:
\be \label{chmasses}
m_e=\sqrt{3} y_e v_d \frac{v}{\Lambda}~~~,~~~~~~~
m_\mu=\sqrt{3} y_\mu v_d \frac{v}{\Lambda}~~~,~~~~~~~
m_\tau=\sqrt{3} y_\tau v_d \frac{v}{\Lambda}~~~.
\ee

In the neutrino sector, after electroweak and $A_4$ symmetry breaking we have Dirac
and Majorana masses:
\be
m_\nu^D=\left(
          \begin{array}{ccc}
            1 & 0 & 0 \\
            0 & 0 & 1 \\
            0 & 1 & 0 \\
          \end{array}
        \right)yv_u\qquad\qquad,~~
M=\left(
\begin{array}{ccc}
A+2 B/3& -B/3& -B/3\\
-B/3& 2B/3& A-B/3\\
-B/3& A-B/3& 2 B/3
\end{array}
\right) u ~~~,
\ee
where 
\be
A\equiv 2 x_A ~~~,~~~~~~~B\equiv 2 x_B \frac{v_S}{u}~~~.
\label{add}
\ee
The mass matrix for light neutrinos is $m_\nu=(m^D_\nu)^T M^{-1} m^D_\nu$ with eigenvalues
\be
m_1=\frac{y^2}{A+B}\frac{v_u^2}{u}~~~,~~~~~~~
m_2=\frac{y^2}{A}\frac{v_u^2}{u}~~~,~~~~~~~
m_3=\frac{y^2}{-A+B}\frac{v_u^2}{u}~~~.
\ee
The mixing matrix is $U_{TB}$, eq. (\ref{2}). 
Both normal and inverted hierarchies 
in the neutrino mass spectrum can be realized. It is interesting that $A_4$ models with the see-saw mechanism typically lead to a light neutrino spectrum which satisfies the sum rule (among complex masses):
\be
\frac{1}{m_3}=\frac{1}{m_1}-\frac{2}{m_2}~~~.\\
\label{sumr}
\ee
A detailed discussion of a spectrum of this type can be found in refs.(\cite{AFmodular,altveram})
Note that in the charged lepton sector the flavour symmetry $A_4$ is broken by $\langle \varphi_T \rangle$ down to
$G_T$. Actually the above mass terms for charged leptons are the most general allowed by the 
symmetry $G_T$. At leading order in $1/\Lambda$, charged lepton masses are diagonal simply because
there is a low-energy $G_T$ symmetry. In the neutrino sector $A_4$ is broken down to $G_S$,
though neutrino masses in this model are not the most general ones allowed by $G_S$. In fact the additional property which is needed, the invariance under $A_{\mu\tau}$, is obtained by stipulating that there are no $A_4$ breaking flavons transforming like 1' and 1".

Recently, in ref. \cite{lam}, the claim was made that, in order to obtain the TB mixing "without fine tuning", the finite group must be $S_4$ or a larger group containing $S_4$. For us this claim is not well grounded being based on an abstract mathematical criterium for a natural model. For us a physical field theory model is natural if the interesting results are obtained from a lagrangian that is the most general given the stated symmetry and the specified representation content for the flavons. For example, we obtain from $A_4$ (which is a subgroup of $S_4$) a natural (in our sense) model for the TB mixing by simply not including symmetry breaking flavons transforming like the 1' and the 1" representations of $A_4$ (a restriction not allowed by the rules specified in ref. \cite{lam} which demand that the symmetry breaking induced by the flavons VEV is the most general). Rather, for naturalness we require that additional physical properties like the hierarchy of charged lepton masses also follow from the assumed symmetry and are not obtained by fine tuning parameters: for this actually $A_4$ can be more effective than $S_4$ because it possesses three different singlet representations 1, 1' and 1''. 

At the next level of approximation each term of the superpotential is corrected by operators of higher dimension whose contributions are suppressed by at least one power of VEV's/$\Lambda$.  The corrections to $w_d$ determine small deviations from the LO VEV alignment configuration. The next to the leading order (NLO) corrections to mass and mixing matrices are obtained by inserting the corrected VEV alignment in the LO operators plus the contribution of the new operators evaluated with the unperturbed VEV's. The final result is \cite{AFmodular} that, when the NLO corrections are included, TB mixing is violated by small terms of the same order for all mixing angles:
\bea
\sin^2\theta_{12}=\frac{1}{3}+{\cal O}(\varepsilon)\nn \\
\sin^2\theta_{23}=\frac{1}{2}+{\cal O}(\varepsilon) \label{corr}\\
\sin\theta_{13}={\cal O}(\varepsilon)\nn 
\eea
As TB mixing is well satisfied by experiment the data require that $\varepsilon < {\cal O}(\lambda_C^2)$.

\section{$A_4$, quarks and GUT's}

Much attention  has been devoted to the question whether models with TB mixing in the neutrino sector can be  suitably extended to also successfully describe the observed pattern of quark mixings and masses and whether this more complete framework can be made compatible with (supersymmetric (SUSY)) SU(5) or SO(10) grand unification. Early attempts of extending models based on $A_4$ to quarks  \cite{ma1.5,AFmodular} and to construct grand unified versions \cite{maGUT} have not been  satisfactory, e.g. do not offer natural mechanisms for mass hierarchies and/or for the vacuum alignment. A direct extension of the $A_4$ model to quarks leads to the identity matrix for $V_{CKM}$ in the lowest approximation, which at first looks promising. But the corrections 
 to it turn out to be strongly constrained by the leptonic sector, because lepton mixings are nearly TB, and, in the simplest models, are proven to be too small to accommodate the observed quark mixing angles \cite{AFmodular}. Also, the quark classification adopted in these models is not compatible with $A_4$ commuting with SU(5) (in ref. \cite{KM} an $A_4$ model compatible with the Pati-Salam group SU(4)$\times$ SU(2)$_L \times$ SU(2)$_R$ has been presented). 
Due to this, larger discrete groups are considered for the description of quarks  and for grand unified versions with approximate TB mixing in the lepton sector. A particularly appealing set of models is based on the discrete group $T'$, the double covering group of $A_4$ \cite{T'0}. In ref. \cite{T'} a viable description was obtained, i.e. in the leptonic sector the predictions of the $A_4$ model are reproduced, while the $T'$ symmetry plays an essential role for reproducing the pattern of quark mixing. But, again, the classification adopted in this model is not compatible with grand unification. Unified models based on the discrete groups $T'$ \cite{CM}, $S_4$ \cite{S4} and $\Delta(27)$  \cite{27} have been 
discussed. Several models using the smallest non-abelian symmetry 
$S_3$ (which is isomorphic to $D_3$) can also be found in the recent literature \cite{S3}.

As a result, the group $A_4$ was considered by most authors to be too
limited to also describe quarks and to lead to a grand unified
description. We have recently shown \cite{afh} that this negative attitude
is not justified and that it is actually possible to construct a
viable model based on $A_4$ which leads to a grand
unified theory (GUT) of quarks and leptons with TB mixing
for leptons. At the same time our model offers an example of an
extra dimensional GUT in which a description of all fermion masses
and mixings is attempted. The model is natural, since most of the
small parameters in the observed pattern of masses and mixings as well
as the necessary vacuum alignment are  justified by the symmetries of
the model. The
formulation of SU(5) in extra dimensions has the usual advantages of
avoiding large Higgs representations to break SU(5) and of solving the
doublet-triplet splitting problem.  A see-saw realization
in terms of an $A_4$ triplet of right-handed neutrinos $N$ ensures the
correct ratio of light neutrino masses with respect to the GUT
scale. In our model extra dimensional effects directly
contribute to determine the flavour pattern, in that the two lightest
tenplets $T_1$ and $T_2$ are in the bulk (with a doubling $T_i$ and
$T'_i$, $i=1,2$ to ensure the correct zero mode spectrum), whereas the
pentaplets $F$ and $T_3$ are on the brane. The hierarchy of quark and
charged lepton masses and of quark mixings is determined by a
combination of extra dimensional suppression factors for the first two
generations and of the U(1) charges, while the neutrino mixing angles
derive from $A_4$. The choice of the transformation properties of the two
 Higgses $H_5$ and $H_{\bar{5}}$ is also crucial. They are chosen to transform 
as two different $A_4$ singlets
$1$ and $1'$. As a consequence, mass terms for the Higgs colour
triplets are  not directly allowed at all orders and their masses are
introduced by orbifolding, \`{a} la Kawamura \cite{5DSU5}. Finally, in this model, proton
decay is dominated by gauge vector boson exchange giving rise to
dimension six operators. Given the relatively large theoretical
 uncertainties, the decay rate is within the present
experimental limits. 
The resulting model is shown to be directly compatible with approximate TB mixing for leptons 
as well as with a realistic pattern of fermion masses and of quark mixings in a SUSY SU(5) 
framework. 

\section{Bimaximal mixing and S4}

We present here the main ideas and results for a model based on $S_4$ that leads to BM mixing in first approximation but the agreement with the data is restored by large NLO corrections that arise from the charged lepton sector. For full details we refer to our paper \cite{S4us}.

The BM mixing matrix is given by:
\begin{equation}
U_{BM}= \left(\matrix{
\dd\frac{1}{\sqrt 2}&\dd-\frac{1}{\sqrt 2}&0\cr
\dd\frac{1}{2}&\dd\frac{1}{2}&-\dd\frac{1}{\sqrt 2}\cr
\dd\frac{1}{2}&\dd\frac{1}{2}&\dd\frac{1}{\sqrt 2}}\right)\;.
\label{21}
\end{equation}
In the BM scheme $\tan^2{\theta_{12}}= 1$, to be compared with the latest experimental
determination:  $\tan^2{\theta_{12}}= 0.45\pm 0.04$ (at $1\sigma$) \cite{data,FogliIndication,MaltoniIndication}, so that a rather large non leading correction is needed, as already mentioned.
In the basis where charged lepton masses are
diagonal, the effective neutrino mass matrix in the BM case is given by
\bea
m_{\nu BM}&=&U_{BM}{\tt diag}(m_1,m_2,m_3)U_{BM}^T\nn\\
&&\nn\\
&=&\left[\frac{m_1}{4}\mathcal{M}_1+\frac{m_2}{4}\mathcal{M}_2+\frac{m_3}{2}\mathcal{M}_3\right]\;.
\label{1k}
\eea
where
\be
\mathcal{M}_1=\left(\matrix{
2&\sqrt 2&\sqrt 2\cr
\sqrt 2&1&1\cr
\sqrt 2&1&1}\right),~~~~~
\mathcal{M}_2=\left(\matrix{
2&-\sqrt 2&-\sqrt 2\cr
-\sqrt 2&1&1\cr
-\sqrt 2&1&1}\right),~~~~~
\mathcal{M}_3=\left(\matrix{
0&0&0\cr
0&1&-1\cr
0&-1&1}\right).
\label{4k}
\ee
The eigenvalues of $m_{\nu}$ are $m_1$, $m_2$, $m_3$ with eigenvectors $(\sqrt{2},1,1)/2$, $(-\sqrt{2},1,1)/2$ and $(0,1,-1)/\sqrt{2}$, respectively.  As we see the most general mass matrix leading to BM mixing is of the form:
\begin{equation}
m_{\nu BM}=\left(\matrix{
x&y&y\cr
y&z&x-z\cr
y&x-z&z}\right)\;,
\label{gl2}
\end{equation}

We now present the argument to show that $S_4$, the permutation group of 4 elements, is a good candidate for a flavour symmetry to realize the BM mixing.  The group $S_4$ has 24 transformations and 5 irreducible representations, which are $3$, $3'$, $2$, $1$ and $1'$.  In terms of two operators $P$ and $R$ satisfying to
\beq
R^4=P^2=(PR)^3=(RP)^3=1\;,
\eeq
all the 24 $S_4$ transformations can be obtained by taking suitable products. Different presentations of the $S_4$ group have been discussed in the recent literature \cite{bmm,Basi} and the choice of generators that we adopt in this paper is related to other existing choices by unitary transformations. Explicit forms of $P$ and $R$ in each of the irreducible representations can be simply obtained. In the representation $1$ we have $R=1$ and $P=1$, while $R=-1$ and $P=-1$ in $1'$. In the representation $2$ we have:
\beq
R=\left(
    \begin{array}{cc}
      1 & 0 \\
      0 & -1 \\
    \end{array}
  \right)\qquad\qquad
P=\frac{1}{2}\left(
    \begin{array}{cc}
      -1 & \sqrt3  \\
      \sqrt3 & 1   \\
    \end{array}
  \right)
  \label{ST2}\;.
\eeq
For the representation $3$, the generators are:
\beq
R=\left(
    \begin{array}{ccc}
      -1 & 0 & 0 \\
      0 & -i & 0 \\
      0 & 0 & i \\
    \end{array}
  \right)\qquad\qquad
P=\left(
    \begin{array}{ccc}
      0 & -\frac{1}{\sqrt{2}} & -\frac{1}{\sqrt{2}}  \\
      -\frac{1}{\sqrt{2}} & \frac{1}{2} & -\frac{1}{2} \\
      -\frac{1}{\sqrt{2}} & -\frac{1}{2} & \frac{1}{2} \\
    \end{array}
  \right)\;.
  \label{matS}
\eeq
In the representation $3'$ the generators $P$ and $R$ are simply opposite in sign with respect to those in the 3.

This description of the group $S_4$ is particularly suitable for our purposes because the general neutrino mass matrix corresponding to the BM mixing, in the basis where charged leptons are diagonal, given by eq. (\ref{gl2}), can be completely characterized by the requirement of being invariant under the action of $P$, eq.~(\ref{matS}), and under the action of $A_{\mu-\tau}$, also a unitary, real and symmetric matrix
defined in eq. (\ref{Amutau}) :
\beq
m_{\nu BM}= P m_{\nu BM}P, ~~~~~~A_{\mu \tau}m_{\nu BM}A_{\mu \tau}~~
\label{invS}
\eeq
As was the case for the $A_4$ models also in this model the invariance under $A_{\mu \tau}$ arises accidentally, as a consequence of the specific field content
and is limited to the contribution of the dominant terms to the neutrino mass matrix. For this reason we do not need to include the $A_{\mu \tau}$ generator in the flavour symmetry group.
Charged leptons must be diagonal in the basis where $m_{\nu}$ has the BM form. A diagonal matrix $m_l^+m_l$  with generic entries is invariant under $R$ given in eq. \ref{matS}:
\beq
m_l^+m_l= R^+ m_l^+m_l R
\eeq
and, conversely, the most general hermitian matrix invariant under $R$ is diagonal (this property remains true also when $R$ is replaced by $\eta R$ where $\eta$ represents an arbitrary phase). By starting from a flavour symmetry group containing $S_4$, we realize a special vacuum alignment such that, at LO, the residual symmetry in the neutrino sector will be that generated by $P$ and $A_{\mu \tau}$, while in the charged lepton sector will be, up to a phase, that generated by $R$. Then, by construction, the LO lepton mixing in this model will be of the BM type. Furthermore, a realistic model will be obtained by adding suitable subleading corrections to this zeroth order approximation.

In the model the 3 generations of left-handed (LH) lepton doublets $l$ and of right-handed (RH) neutrinos $\nu^c$ to two triplets $3$, while the RH charged leptons $e^c$, $\mu^c$ and $\tau^c$ transform as $1$, $1'$ and $1$, respectively. The $S_4$ symmetry is then broken by suitable triplet flavons. All the flavon fields are singlets under the Standard Model gauge group.  Additional symmetries are needed, as usual, to prevent unwanted couplings and to obtain a natural hierarchy among $m_e$, $m_\mu$ and $m_\tau$. In our model, the complete flavour symmetry is $S_4\times Z_4\times U(1)_{FN}$.  A flavon $\theta$, carrying a negative unit of the $U(1)_{FN}$ charge F, acquires a vacuum expectation value (VEV) and breaks $U(1)_{FN}$. In view of a possible GUT extension of the model at a later stage, we adopted a supersymmetric context, so that two Higgs doublets $h_{u,d}$, invariant under $S_4$, are present in the model as well as the $U(1)_R$ symmetry related to R-parity and the driving fields in the flavon superpotential. Supersymmetry also helps producing and maintaining the hierarchy $\langle h_{u,d}\rangle=v_{u,d}\ll \Lambda$ where $\Lambda$ is the cut-off scale of the theory.

The fields in the model and their classification under the symmetry are summarized in Table \ref{table:TransformationsS}. The fields  $\psi_l^0$, $\chi_l^0$ , $\xi_\nu^0$ and $\phi_\nu^0$ are the driving fields. \begin{table}[h]
\begin{center}
\begin{tabular}{|c||c|c|c|c|c|c||c||c|c|c|c||c|c|c|c|}
  \hline
  &&&&&&&&&&&&&&&\\[-0,3cm]
  & $l$ & $e^c$ & $\mu^c$ & $\tau^c$ & $\nu^c$ & $h_{u,d}$ & $\theta$ & $\phi_l$ & $\chi_l$ & $\psi_l^0$ & $\chi_l^0$ & $\xi_\nu$ &$\phi_\nu$ & $\xi_\nu^0$ & $\phi_\nu^0$ \\
  &&&&&&&&&&&&&&&\\[-0,3cm]
  \hline
  &&&&&&&&&&&&&&&\\[-0,3cm]
  $S_4$ & 3 & 1 & $1^\prime$ & 1 & 3 & 1 & 1 & 3 & $3^\prime$ & 2 & $3'$ & 1 & 3 & 1 & 3  \\
  &&&&&&&&&&&&&&&\\[-0,3cm]
  $Z_4$ & 1 & -1 & -i & -i & 1 & 1 & 1 & i & i & -1 & -1 & 1 & 1 & 1 & 1 \\
  &&&&&&&&&&&&&&&\\[-0,3cm]
  $U(1)_{FN}$ & 0 & 2 & 1 & 0 & 0 & 0 & -1 & 0 & 0 & 0 & 0 & 0 & 0 & 0 & 0  \\
  &&&&&&&&&&&&&&&\\[-0,3cm]
  $U(1)_R$ & 1 & 1 & 1 & 1 & 1 & 1 & 0 & 0 & 0 & 2 & 2 & 0 & 0 & 2 & 2  \\
  \hline
  \end{tabular}
\end{center}
\begin{center}
\begin{minipage}[t]{12cm}
\caption[]{\label{table:TransformationsS}Transformation properties of all the fields.}
\end{minipage}
\end{center}
\end{table}
The complete superpotential can be written as $w=w_l+w_\nu+w_d$.
The $w_d$ term is responsible for the alignment and will not be discussed here. The terms $w_l$ and $w_\nu$
determine the lepton mass matrices (we indicate with $(\ldots)$ the singlet 1, with $(\ldots)^\prime$ the singlet $1^\prime$ and with $(\ldots)_V$ ($V=2,\,3,\,3'$) the representation V)
\bea
w_l\;=&&\frac{y_e^{(1)}}{\La^2}\frac{\theta^2}{\La^2}e^c(l\phi_l\phi_l)+ \frac{y_e^{(2)}}{\La^2}\frac{\theta^2}{\La^2}e^c(l\chi_l\chi_l)+ \frac{y_e^{(3)}}{\La^2}\frac{\theta^2}{\La^2}e^c(l\phi_l\chi_l)+\nn\\
&+&\frac{y_\mu}{\La}\frac{\theta}{\La}\mu^c(l\chi_l)^\prime+\frac{y_\tau}{\La}\tau^c(l\phi_l)+\dots
\label{wl}\\
\nn\\
w_\nu\;=&&y(\nu^cl)+M \Lambda (\nu^c\nu^c)+a(\nu^c\nu^c\xi_\nu)+b(\nu^c\nu^c\phi_\nu)+\dots\\
\label{wd}\,
\eea
Again, to keep our formulae compact, we omit to write the Higgs fields
$h_{u,d}$.  For instance
$y_\tau \tau^c(l\phi_l)/\La$ stands for $y_\tau \tau^c(l\phi_l)h_d/\La$,
$y(\nu^cl)$ stands for $y(\nu^cl) h_u$. The powers of the cutoff $\La$ also take  into account the presence of the omitted Higgs fields. Note that the parameters $M$, $M_\phi$, $M_\xi$ and $M'_\xi$ defined above are dimensionless.
In the above expression for the superpotential $w$, only the lowest order operators
in an expansion in powers of $1/\Lambda$ are explicitly shown. Dots stand for higher
dimensional operators that will be discussed later on. The stated symmetries ensure that, for the leading terms, the flavons that appear in $w_l$ cannot contribute to $w_\nu$ and viceversa.

The potential corresponding to $w_d$ possesses an isolated minimum for the following VEV configuration:
\beq
\dd\frac{\mean{\phi_l}}{\La}=\left(
                     \begin{array}{c}
                       0 \\
                       1 \\
                       0 \\
                     \end{array}
                   \right)A\qquad\qquad
                   \dd\frac{\mean{\chi_l}}{\La}=\left(
                     \begin{array}{c}
                       0 \\
                       0 \\
                       1 \\
                     \end{array}
                   \right)B
\label{vev:charged:best}
\eeq
\beq
\hspace{-1.5cm}
\dd\frac{\mean{\phi_\nu}}{\La}=\left(
                     \begin{array}{c}
                       0 \\
                       1 \\
                       -1 \\
                     \end{array}
                   \right)C\qquad\quad
\dd\frac{\mean{\xi_\nu}}{\La}=D
\label{vev:neutrinos}
\eeq
where the factors  $A$, $B$, $C$, $D$ should obey to the relations:
\bea
&\sqrt{3}f_1A^2+\sqrt{3}f_2B^2+f_3AB=0
\label{AB}\\
\nn\\
&D=-\dd\frac{M_\phi}{g_2}\qquad\qquad C^2=\dd\frac{g_2^2M_\xi^2+g_3M_\phi^2-g_2M_\phi M'_\xi}{2 g_2^2g_4}
\label{CD}\;.
\eea
Similarly, the Froggatt-Nielsen flavon $\theta$ gets a VEV, determined by the D-term associated to the local $U(1)_{FN}$ symmetry, and it is denoted by
\beq
\frac{\mean{\theta}}{\La}= t\;.
\label{deft}
\eeq

With this VEV's configuration, the charged lepton mass matrix is diagonal
\beq
m_l=\left(
         \begin{array}{ccc}
           (y_e^{(1)}B^2-y_e^{(2)}A^2+y_e^{(3)}AB)t^2 & 0 & 0 \\
           0 & y_\mu Bt & 0 \\
           0 & 0 & y_\tau A \\
         \end{array}
       \right) v_d
\eeq
so that at LO there is no contribution to the $U_{PMNS}$ mixing matrix from the diagonalization of charged lepton masses.
In the neutrino sector for the Dirac and RH Majorana matrices we have
\beq
m_\nu^D=\left(
          \begin{array}{ccc}
            1 & 0 & 0 \\
            0 & 0 & 1 \\
            0 & 1 & 0 \\
          \end{array}
        \right)yv_u\qquad\qquad
M_N=\left(
              \begin{array}{ccc}
                2M+2aD & -2bC & -2bC \\
                -2bC & 0 & 2M+2aD \\
                -2bC & 2M+2aD & 0 \\
              \end{array}
            \right)\Lambda\;.
\label{Feq:RHnu:masses}
\eeq
The matrix $M_N$ can be diagonalized by the BM mixing matrix $U_{BM}$, which represents the full lepton mixing at the LO, and the eigenvalues are
\beq
M_1=2|M+aD-\sqrt{2}bC|\Lambda\qquad M_2=2|M+aD+\sqrt{2}bC|\Lambda\qquad M_3=2|M+aD|\Lambda\;.
\eeq
After see-saw, since the Dirac neutrino mass matrix commutes with $M_N$ and its square is a matrix proportional to unity,
the light neutrino Majorana mass matrix, given by the see-saw relation \mbox{$m_\nu=(m_\nu^D)^TM_N^{-1}m_\nu^D$}, is also diagonalized by the BM mixing matrix and the eigenvalues are
\beq
|m_1|=\frac{|y^2|v_u^2}{2|M+aD-\sqrt{2}bC|}\dd\frac{1}{\Lambda}\qquad
|m_2|=\frac{|y^2|v_u^2}{2|M+aD+\sqrt{2}bC|}\dd\frac{1}{\Lambda}\qquad
|m_3|=\frac{|y^2|v_u^2}{2|M+aD|}\dd\frac{1}{\Lambda}\;.
\label{spec}
\eeq
The light neutrino mass matrix depends on only 2 effective parameters, at LO, indeed the terms $M$ and $aD$ enter the mass matrix in the combination $F\equiv M+a D$. The coefficients $y_e^{(i)}$, $y_\mu$, $y_\tau$, $y$, $a$ and $b$ are all expected to be of $\mathcal{O}(1)$. A priori $M$ could be of $\mathcal{O}(1)$, corresponding to a RH neutrino Majorana mass of $\mathcal{O}(\Lambda)$, but, actually,  it must be of the same order as $C$ and $D$. In the context of a grand unified theory this would correspond to
the requirement that $M$ is of $\mathcal{O}(M_{GUT})$ rather than of $\mathcal{O}(M_{Planck})$.

We expect a common order of magnitude for the VEV's (scaled by the cutoff $\Lambda$):
\beq
A \sim B \sim v\;,~~~~~~~~~~C \sim D \sim v'\;.
\eeq
However, due to the different minimization conditions that determine $(A,B)$ and $(C,D)$, we may tolerate a moderate hierarchy
between $v$ and $v'$. Similarly the order of magnitude of $t$ is in principle unrelated to those of $v$ and $v'$.
It is possible to estimate the values of $v$ and $t$ by looking at the mass ratios of charged leptons:
\bac{ll}
\left(\dd\frac{m_\mu}{m_\tau}\right)_{exp}\simeq0.06&\qquad\dd\frac{m_\mu}{m_\tau} \sim t\\
\\[-0,3cm]
\left(\dd\frac{m_e}{m_\mu}\right)_{exp}\simeq0.005&\qquad\dd\frac{m_e}{m_\mu} \sim vt\\
\\[-0,3cm]
\left(\dd\frac{m_e}{m_\tau}\right)_{exp}\simeq0.0003&\qquad\dd\frac{m_e}{m_\tau} \sim vt^2\;.
\eac
In order to fit these relations, approximately we must have $t \sim 0.06$ and $v \sim 0.08$ (modulo coefficients of $\mathcal{O}(1)$).\\

So far we have shown that, at LO, we have diagonal and hierarchical charged leptons together with the exact BM mixing for neutrinos. It is clear that substantial NLO corrections are needed to bring the model to agree with the data on $\theta_{12}$. A crucial feature of our model is that the neutrino sector flavons  $\phi_\nu$ and $\xi_\nu$ are invariant under $Z_4$ which is not the case for the charged lepton sector flavons $\phi_l$ and $\chi_l$. The consequence is that $\phi_\nu$ and $\xi_\nu$  can contribute at NLO to the corrections in the charged lepton sector, while at NLO $\phi_l$ and $\chi_l$ cannot modify the neutrino sector couplings. As a results the dominant genuine corrections to the BM mixing only occur  at NLO through the diagonalization of the charged leptons. 
In fact, at NLO the neutrino mass matrix is still diagonalized by $U_{BM}$ but the mass matrix of charged leptons is no more diagonal. Including these additional terms from the diagonalization of charged leptons the $U_{PMNS}$ matrix can be written as
\beq
U_{PMNS}=U_l^\dag U_{BM}\;,
\eeq
and therefore the corrections from $U_l$ affect the neutrino mixing angles at NLO according to
\bac{l}
\sin^2\theta_{12}=\dd\frac{1}{2}-\frac{1}{\sqrt{2}}(V_{12}+V_{13})v'\\[0.2cm]
\sin^2\theta_{23}=\dd\frac{1}{2}\\[0.2cm]
\sin\theta_{13}=\dd\frac{1}{\sqrt{2}}(V_{12}-V_{13})v'\;.
\label{sinNLO}
\eac

By comparing these expressions with the current experimental values of the mixing angles in table \ref{table:OscillationData}, we see that, to correctly reproduce $\theta_{12}$ we need a parameter $v'$ of the order of the Cabibbo angle $\lambda_C$. Moreover, barring cancellations of/among some the $V_{ij}$ coefficients, also $\theta_{13}$ is corrected by a similar amount, while $\theta_{23}$ is unaffected at the NLO. A salient feature of our model is that, at NLO accuracy, the large corrections of $\mathcal{O}(\lambda_C)$ only apply to $\theta_{12}$ and $\theta_{13}$ while $\theta_{23}$ is unchanged at this order. As a correction of $\mathcal{O}(\lambda_C)$ to $\theta_{23}$ is hardly compatible with the present data (see table \ref{table:OscillationData}) this feature is very crucial for the phenomenological success of our model. It is easy to see that this essential property depends on the selection in the neutrino sector of flavons $\xi_\nu$ and $\phi_\nu$ that transform as 1 and 3 of $S_4$, respectively. We have checked that if, for example, the singlet $\xi_\nu$ is replaced by a doublet $\psi_\nu$ (and correspondingly the singlet driving field  $\xi_\nu^0$ is replaced by a doublet $\psi_\nu^0$), all other quantum numbers being the same, one can construct a variant of the model along similar lines, but in this case all the 3 mixing angles are corrected by terms of the same order. This confirms that a  particular set of $S_4$ breaking flavons is needed in order to preserve $\theta_{23}$ from taking as large corrections as the other two mixing angles.

All this discussion applies at the NLO  and we expect that at the NNLO the value of $\theta_{23}$ will eventually be modified with deviations of about $\mathcal{O}(\lambda^2_C)$. The next generation of experiments, in particular those exploiting a high intensity neutrino beam, will probably reduce the experimental error on $\theta_{23}$ and the sensitivity on $\theta_{13}$ to few degrees. All quantitative estimates are clearly affected by large uncertainties due to the presence of unknown parameters of order one, but in our model a value of $\theta_{13}$ much smaller than the present upper bound would be unnatural. If in the forthcoming generation of experiments no significant deviations from zero of $\theta_{13}$ will be detected, our construction will be strongly disfavoured.

\section{Conclusion}

The present situation is that the data on neutrino mixing are in agreement, within 1-$\sigma$ or so, with TB mixing. If this indication is taken seriously than models that can reproduce TB mixing with good approximation are of prominent interest. Among those a particularly simple and, by now, well studied class of models are those based on the $A_4$ discrete group. These models are particularly successful and attractive in the lepton sector. Their extension to quarks and their embedding in a Grand Unified picture, although possible (several examples exist)
is not straightforward: in particular the quark mixings and hierarchies are typically determined by additional mechanisms other than the $A_4$ symmetry. Alternatively we can assume that the agreement with TB mixing is accidental. There are in fact models where the TB mixing pattern is not implied but that fit the data (in many cases different values of the mixing angles could also be reproduced in these models by changing the values of the parameters). An interesting possibility, suggested by the concept of "weak complementarity" (defined in the Introduction), is that the relevant first approximation is BM mixing corrected by relatively large terms arising from the diagonalization of charged leptons. We have developed a model based on the discrete group $S_4$ where this scheme is realized. A signal for this kind of model would be a value of  $\theta_{13}$ near the present upper bound, while in most $A_4$ models for approximate TB mixing we expect  $\theta_{13}\sim o(\lambda_C^2)$. 

\section{Acknowledgments}
It is a very pleasant duty for me to most warmly thank Professor Milla Baldo-Ceolin for her kind invitation and for the great hospitality offered to all of us in Venice.

\end{document}